# Observation of the Efimov state of the helium trimer

**Authors:** Maksim Kunitski[1]*, Stefan Zeller[1], Jörg Voigtsberger[1], Anton Kalinin[1], Lothar Ph. H. Schmidt[1], Markus Schöffler[1], Achim Czasch[1], Wieland Schöllkopf[2], Robert E. Grisenti[3], Till Jahnke[1], Dörte Blume[4] and Reinhard Dörner[1]*

**Affiliations:**

[1]Institut für Kernphysik, Goethe-Universität Frankfurt am Main, Max-von-Laue-Straße 1, 60438 Frankfurt am Main, Germany.

[2]Department of Molecular Physics, Fritz-Haber-Institut, Faradayweg 4-6, 14195 Berlin, Germany

[3]GSI - Helmholtz Center for Heavy Ion Research, Planckstraße 1, 64291 Darmstadt, Germany

[4]Department of Physics and Astronomy, Washington State University, Pullman, WA 99164-2814, USA

*Correspondence to: kunitski@atom.uni-frankfurt.de, doerner@atom.uni-frankfurt.de

**Abstract**: Quantum theory dictates that upon weakening the two-body interaction in a three-body system, an infinite number of three-body bound states of a huge spatial extent emerge just before these three-body states become unbound. Three helium atoms have been predicted to form a molecular system that manifests this peculiarity under natural conditions without artificial tuning of the attraction between particles by an external field. Here we report experimental observation of this long predicted but experimentally elusive Efimov state of $^4\text{He}_3$ by means of Coulomb explosion imaging. We show spatial images of an Efimov state, confirming the predicted size and a typical structure where two atoms are close to each other while the third is far away.

**One Sentence Summary:** We report experimental discovery of a gigantic molecule that consists of three helium atoms and is bound solely by a universal feature of quantum mechanics called "Efimov effect".

Ever since the early days of celestial mechanics, the three-body problem posed a major challenge to physicists. In the early 20th century the failure of finding a stable solution for the classical helium atom (2 electrons and a nucleus) heralded the demise of Niels Bohr's program of semi-classical atomic physics (*1*). Quantum mechanics then added yet another surprising twist to the three-body problem when in 1970 Vitaly Efimov predicted the appearance of an infinite series of stable three-body states of enormous spatial extents (*2*). These Efimov states are predicted to exist for short-range interactions like the van der Waals force between atoms or the strong force between nucleons. When the potential becomes so shallow that the last two-body bound state is at the verge of becoming unbound or is unbound, then three particles stick together to form Efimov states. Intriguingly, this three-body behavior does not depend on the details of the underlying two-body interactions. This makes the Efimov effect a universal phenomenon, with important applications in particle, nuclear (*3*, *4*), atomic (*4*), condensed matter (*5*) and biological physics (*6*).

Figure 1 summarizes two facets of Efimov's prediction, namely the energy spectrum and the structure of an Efimov state. Figure 1A shows how the two- and three-body binding energies (the binding energy of an atomic cluster is defined as the energy needed to separate all constituents of the cluster to infinite distances) change as the depth of the two-body potential is increased. As indicated by the arrow above Figure 1A, the depth of the two-body potential increases along the horizontal axis. As the depth increases, the s-wave scattering length *a* changes from negative values to infinitely large values to positive values. Negative *a* values correspond to the domain where shallow two-body bound states do not exist. For positive *a*, a shallow two-body bound state, the dimer (see the blue solid line), exists. Bound three-body states (called trimers) exist in the green-shaded area. The extremely weakly-bound three-body states close to threshold (see the solid red line labeled "1st ES" and the dashed black line labeled "2nd ES" in Figure 1A) are Efimov states, which have been predicted to possess remarkable characteristics that are intricately related to the discrete scale invariance of the underlying three-body Hamiltonian. Key characteristics of Efimov states are their unusual extent and structure. Figure 1B shows the calculated structure of the state labeled "1st ES" for $\text{sign}(a)|a|^{-1/2}=0$, i.e., for the ideal and universal case where the two-body scattering length is infinitely large and the dimer binding energy is equal to zero. For comparison, the size and shape of the ground state trimer (labeled "GS" in Figure 1A) are depicted in Figure 1C. The hypothetical ideal Efimov trimer extends out to 300 Angstrom (1 Angstrom (Å) = 0.1 nm), i.e., the Efimov trimer is about 100 times larger than a typical chemically-bound triatomic molecule. Moreover, the ideal Efimov state is highly diffuse and does not display a predominantly equilateral triangular or linear shape.

Despite their relevance across different subfields of physics, these spatially extended and weakly-bound trimer states have proven extremely challenging to prepare and detect, and experimental evidence for the Efimov effect was reported only in 2006 (*7*) (i.e., 36 years after its theoretical prediction), simulating a great deal of continued experimental activity. Experimental signatures to date have come from loss measurements or spectroscopy on trapped cold atom systems. In experiments of this type, the two-body interaction is tuned in the vicinity of a Feshbach resonance through the application of an external magnetic field. Signatures of the Efimov effect are then obtained by monitoring the atom loss from the trap due to the formation of Efimov trimers at scattering lengths for which the trimer energy coincides with that of three free atoms (these scattering lengths are marked by stars in Figure 1A). The most direct probe of Efimov trimers to date comes from radio frequency spectroscopy on an ultracold three-component lithium gas, which yielded the binding energies of an Efimov trimer for different two-body interaction strengths (*8*, *9*). An experimental exploration of the size and shape of Efimov states requires a setup where the trimer is sufficiently long lived and can be imaged selectively. These two demands prove challenging for cold atom experiments. The experiments reported in this work circumvent these challenges by working with a different species, namely $^4$He, and a completely different experimental approach.

The helium trimer is a paradigmatic molecular system that is believed to support an Efimov state. In fact, it was already predicted in 1977 to be a prime candidate to study Efimov physics (*10*). Theoretical calculations based on the currently most accurate He-He potential (*11*) predict two bound states for the helium trimer, the ground and excited states, with binding energies of 131.84 mK and 2.6502 mK, respectively (*12*). These states occur naturally along the vertical dashed line in Figure 1A at a scattering length of *a*=90.4 Å (*11*). The excited state has one node in the hyperradial coordinate, indicating a vibrational excitation that is reminiscent of a breathing mode in classical tri-atomic molecules. Our calculations, shown by the green and red lines in Figure 1A, conclude - in agreement with earlier works (*4*, *10*, *13*–*16*) – that an artificial strengthening of the pair interaction (see the right pointing arrow above Figure 1A) renders the excited state of the helium trimer less strongly bound with respect to the dimer, thus confirming the Efimov character of the trimer state. This character is further supported by the appearance of a second Efimov state upon an artificial weakening of the true two-body He-He potential (dashed black line labeled "2nd ES" in Figure 1A).

The energy ratio of two neighboring Efimov states for infinitely large s-wave scattering length is $22.7^2$. Correspondingly, the excited state is about 22.7 larger than the ground state (see Figures 1B and 1C). The energy ratio decreases when the scattering length takes finite positive values. For the scattering length corresponding to helium, Efimov's radial law predicts an energy ratio of $5.0^2$. The energy ratio of the ground and excited state of the true helium trimers is larger than predicted by Efimov's theory, namely $7.0^2$, owing to the fact that the ground state of the helium trimer is not an Efimov state.

Because of the very low binding energy, the $^4$He trimer lacks rotational states, which implies that common structural experimental tool such as rotational spectroscopy cannot be applied. The $^4$He trimer ground state was observed experimentally in 1996 by Schöllkopf and Toennies (*17*) utilizing matter wave diffraction of helium clusters from a transmission grating. Concerted experimental efforts, however, did not provide any evidence for the existence of the excited state of the $^4$He trimer (*18*, *19*).

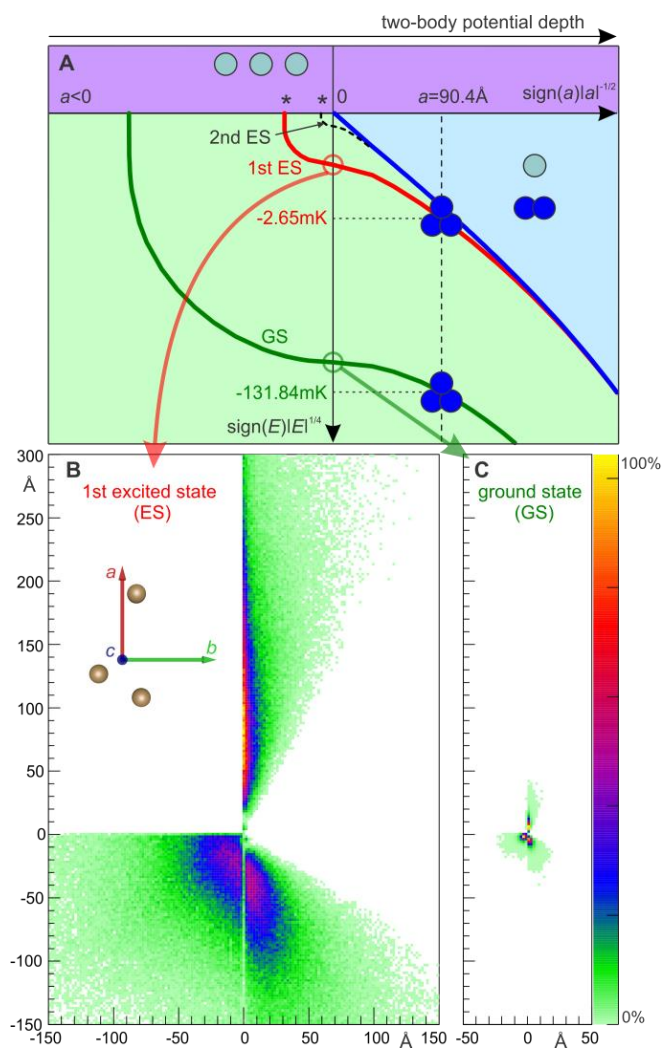

**Fig. 1. Theoretical basis for the Efimov state of the helium trimer. A:** Dependence of the binding energies (sign($E$)·|$E$|$^{1/4}$) of the ground (GS) and first excited (1st ES) states of the helium trimer on the scattering length (sign($a$)·|$a$|$^{-1/2}$) calculated by artificially scaling the He-He potential. The blue line on the positive scattering length side shows the binding energy of the helium dimer. The vertical dashed line corresponds to the naturally occurring $^4$He system with a two-body scattering length of 90.4 Å (*11*). **B and C:** Theoretical structures of the excited and ground state of the hypothetical helium trimer corresponding to the scaled He-He potential with an infinite scattering length. The structures are plotted in the principal axis frame (*abc*) as shown in the inset. Namely, the center-of-mass of the trimer was shifted to the origin and the structures were rotated such that the principal axis with the smallest moment of inertia (shown by the red vector *a* in the inset) lies along the y-axis. Additionally, if required, the structure was mirrored with respect to the x- or y-axis in order to get one helium atom in the first quadrant and the other two in the third and forth quadrants, so that the second quadrant is always empty. The linear color scale encodes the number of entries.

In the present work we report experimental observation of the Efimov state of the helium trimer; the creation of this stable state in a "natural" field-free environment allows us to directly investigate the structural aspects of Efimov physics and take real space images of the square of the wave function of an Efimov state using Coulomb explosion imaging. The obtained experimental distributions are in good agreement with those obtained from full first principles quantum mechanical calculations (*20*) that utilize the currently most accurate He-He potential published by Cencek et al. (*11*).

The He clusters were prepared in a molecular beam under supersonic expansion of gaseous helium at a temperature of 8 K through a 5 µm nozzle. The cluster yields were tuned by varying the nozzle back pressure. Helium trimers were selected from the molecular beam by means of matter wave diffraction (*17*). The selection removed an overwhelming fraction of helium monomers that dominates the molecular beam under all expansion conditions. All three atoms of a trimer were then singly ionized by a strong ultra-short laser field (30 fs, 780 nm, >3·10$^{15}$ W/cm$^2$). Since the ionization process is essentially instantaneous, the quantum mechanical probability distributions of the neutral trimers provide the initial configurations for the subsequent Coulomb explosion (*21*) of the triply charged trimers. The momenta acquired during the explosion of the ions were measured by cold target recoil ion momentum spectroscopy (COLTRIMS) (*22, 23*). From these momentum vectors the initial spatial geometry of the three charged fragments at the instant of ionization was reconstructed using Newton's equation of motion (for details see the supplementary materials). A simple global observable related to the structure is the total kinetic energy of all three ions (kinetic energy release, KER). In the Coulomb explosion the total potential energy of the three charges with interparticle distances $R_{ij}$ is converted into KER (in atomic units):

$$\text{KER} = 1/R_{12} + 1/R_{13} + 1/R_{23} \quad (1)$$

The measured KER distributions corresponding to the He trimer at two different nozzle back pressures are depicted in Figure 2A. At a pressure of 1.7 bar (Figure 2A, blue curve) only one peak with a maximum at 5 eV is observed. This peak corresponds to the ground state of the helium trimer with an average He-He distance of 10.4 Å (*24*). At a lower nozzle pressure of 330 mbar (Figure 2A, black curve) an additional peak with a maximum at 0.57 eV emerges. This low-energy feature corresponds to structures that are about eight times larger than those of the ground state. Such large spatial extent (≈80 Å) has been predicted for the excited state of the $^4$He trimer (*12, 25, 26*). Indeed, the KER distribution (Figure 2A, violet curve) calculated from the full quantum mechanical probability distribution of the excited helium trimer using eq 1 resembles the experimental observation (difference spectrum in red, Figure 2A) very closely.

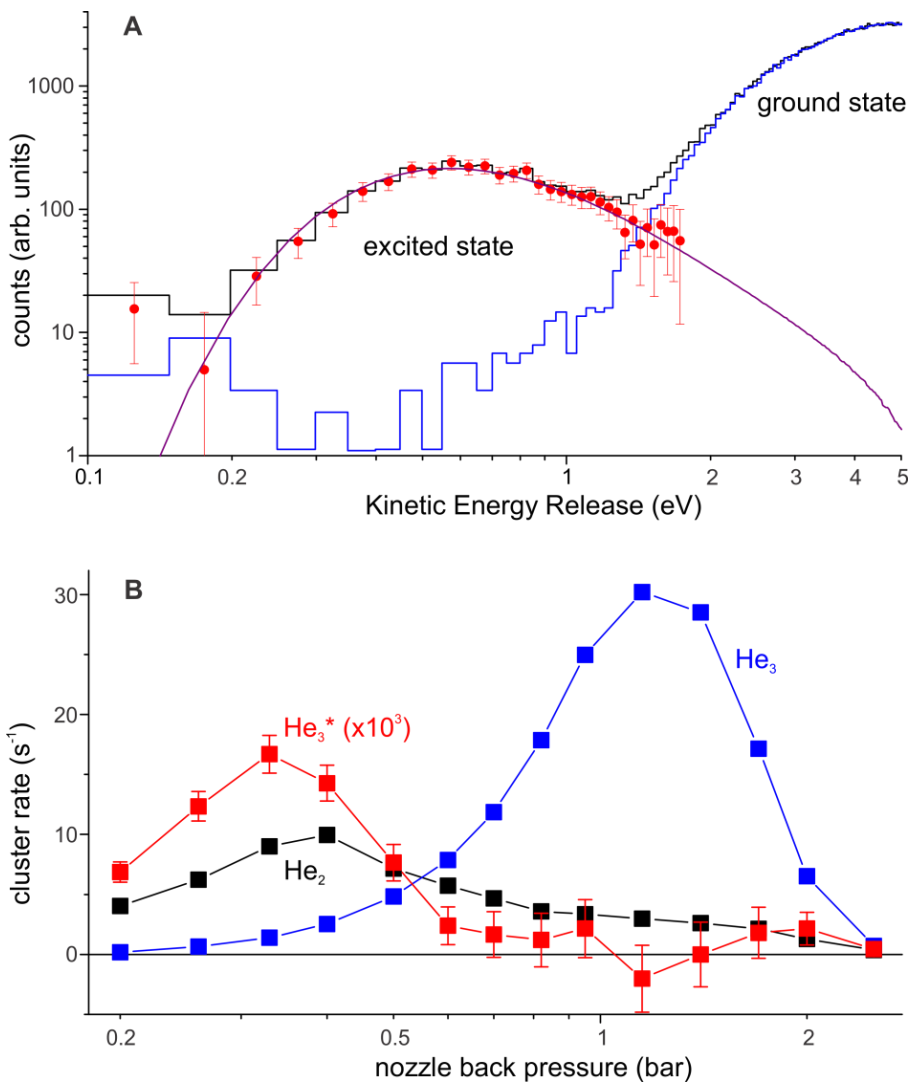

**Fig. 2. Experimental observation of the He$_3$ excited state. A:** The kinetic energy release (KER) distributions of Coulomb exploded $^4$He trimers. The experimental distributions correspond to the mixture of the ground and excited states (black, expansion conditions: 8 K, 330 mbar) and to the ground state only (blue, expansion conditions: 8 K, 1.7 bar). The difference spectrum is shown in red with error bars (correspond to a confident interval of 95%). The ground state only distribution (blue) is normalized such that it agrees with the mixture distribution (black) for KER around 3-5 eV. The theoretical KER distribution for the excited state of the $^4$He trimer, obtained from our full quantum mechanical calculation, is shown in violet. Note the logarithmic scale on both axes. **B:** Dependence of the $^4$He cluster rates on the back pressure at a temperature of 8 K for a nozzle with a 5 μm orifice. The very low rate of the He$_3$ excited state (red) is scaled by a factor of $10^3$. The background caused by ground state structures has been subtracted from the excited state rate. The error bars correspond to a confident interval of 68%. The rates for the He$_3$ ground state and He$_2$ are shown in blue and black, respectively.

As Figure 2A shows, the yield of the $^4$He$_3$ excited state is sensitive to the expansion conditions. The detailed analysis of the pressure dependence shows that the maximum rate of the excited state of the $^4$He trimer is achieved not at pressures with the highest yield of the ground state of the trimer but rather at pressures that favor dimer formation (Figure 2B). This might be an indication that two He dimers are required for the formation of the excited He trimer during the supersonic expansion; the formation mechanism of the ground state trimer is seeded primarily by collisions between one helium dimer and two helium monomers (*27*). Another interpretation of the observed pressure dependence of the excited state yield might be an increased collision-induced break-up rate of excited trimers in an expansion at higher pressures due to increased translational temperatures (*27*). The highest ratio of the excited to the ground state He trimer populations of about 4% was found at the lowest pressure used in the experiment, namely at 200 mbar. At a pressure with the maximum yield of the trimer ground state we could not detect any contribution of the excited state. This very weak relative yield of the excited state explains why it was not observed in the experiment of Toennies and coworkers (*19*) with an estimated detection limit of 6 %.

Figure 2 establishes unambiguously that the helium trimer excited state is stable and can be prepared reliably in an experiment. In order to deduce quantitative information about the structural properties of the excited He trimer from the measured momenta we have used classical mechanics to invert the Coulomb explosion (see (*24*) and the supplementary materials for details). Unfortunately, there is an ambiguity of momentum-to-structure relation in the small region of the structural space. This results in reconstruction of some number of irrelevant geometries. In order to overcome this issue, the irrelevant structures were filtered out during reconstruction (for details see the supplementary materials).

Figure 3 shows the reconstructed pair distance distributions for the excited state (red and black) and the ground state (blue) of He$_3$. The red and black distributions differ in how the excited state structures are separated from the ground state structures. The separation is necessary because the wave functions of both states overlap spatially in the range of lower pair distances and the ground state always dominates in the experiment. By applying a filter in the momentum space (for details see the supplementary materials) we were able to choose momenta that mainly relate to the excited state of He$_3$ (Figure 3, black). However some structures of the excited state have been cut by the filter resulting in the discrepancy below a pair distance of 50 Å. An alternative way of obtaining the excited state distribution is subtraction of the reconstructed distribution of the ground state from the distribution of the mixture of the ground and excited states (Figure 3, red). This approach still fails in an accurate reproduction of the distribution in the lower pair distance range, because of the overwhelming amount of the ground state, whose distribution spreads up to 50 Å (Figure 3, blue). At larger distances (>100 Å), however, both experimental pair distance distributions of the excited state are nearly identical and match the theoretical distribution (Figure 3, violet) very well.

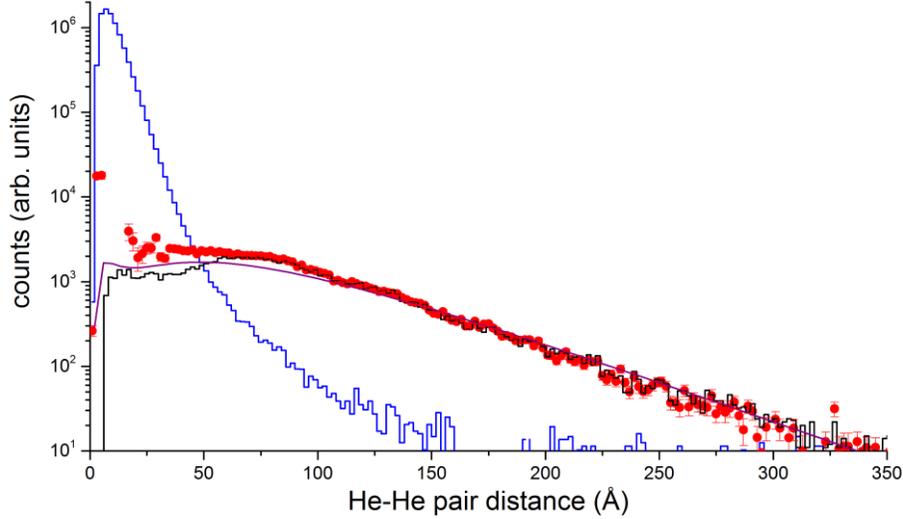

**Fig. 3. Pair distance distributions $P_{pair}(R)$ of the He$_3$ excited state.** The red circles represent the difference between the mixture of the excited and ground state distribution (measured at a nozzle pressure of 330 mbar and a temperature of 8 K) and the ground state only distribution (1.7 bar, 8 K, blue line). The error bars correspond to a confident interval of 68%. The ground state distribution was normalized to the quantity of ground state structures under conditions where the excited state was measured (300 mbar, 8 K). The black histogram corresponds to the distribution that was obtained from the measured momenta of the ground and excited state mixture by filtering out the structures with higher KERs (for details see the supplementary materials). Experimental distributions have been reconstructed from the measured momenta using Newtonian mechanics to invert the Coulomb explosion. The theoretical pair distance distribution of the excited helium trimer is shown in purple.

The measured pair distribution (Figure 3) can be used to extract the binding energy of the excited state. The excited He$_3$ exists mainly in the classically forbidden region well outside the two-body interaction potential well. Therefore, the asymptotic part of the pair distance distribution $P_{pair}(R)$ can be approximated by an exponential decay function (*12*):

$$P_{pair}(R) \propto e^{-2\sqrt{2\mu\Delta B/\hbar^2}R} \qquad (2),$$

where $R$ denotes the He-He distance, $\mu = 2 \cdot m_{He}/3$ is the reduced mass of the helium trimer and $\Delta B$ is the difference between the binding energies of the excited trimer state and the dimer ground state. Fitting the falling edge of the reconstructed pair distance distribution (black curve, Figure 3) by the function (2) yields $\Delta B$ of 0.98±0.2 mK (for details see the supplementary materials). Given a theoretical prediction for the dimer binding energy of 1.62 mK (*11*), we obtain a binding energy of the trimer excited state of 2.6±0.2 mK. This is in excellent agreement with our theoretical value of 2.65 mK as well as with a theoretical value of 2.6502 mK from Ref. (*12*).

Having established the spatial extent of the He$_3$ Efimov state, we now discuss its geometric shape revealed by the structural plots shown in Figure 4. The plots in the upper row (A-C) of Figure 4 are generated in the center-of-mass coordinate frame with the principal axis of the smallest moment of inertia chosen to lie along the y-axis as proposed by Nielsen et al. (*25*). In the plots of the lower row (D-F) of Figure 4, the interparticle distances of each reconstructed geometry were initially normalized to the largest of the three interparticle distances. Subsequently, the two atoms with the largest pair distance were placed at positions (-0.5,0) and (0.5,0) and the position of the third atom was plotted. In these plots the equilateral triangle corresponds to the (x,y)=(0,sqrt(3)/2) and linear configurations have y=0. The structure in which two particles are close to each other, with the third particle being far away, corresponds to (x,y) being close to (-0.5,0) or (0.5,0).

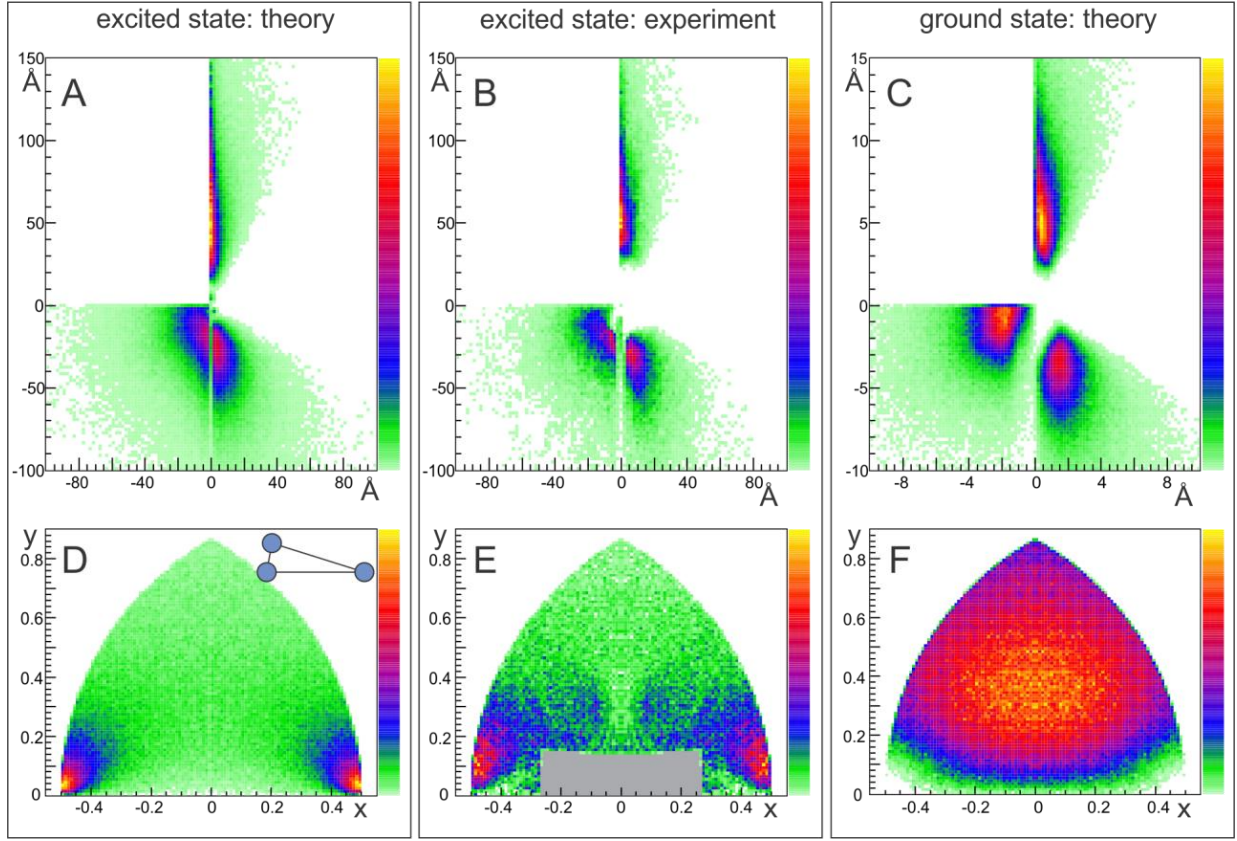

**Fig. 4. Structures of the helium trimer. A, B respectively show the theoretical and experimental excited state structure; C shows the theoretical ground state. Note the different scale for the ground state structure (C). For the plots the center-of-mass of the trimer was shifted to the origin. The structures were rotated such that the principal axis with the smallest moment of inertia lies along the y-axis. Additionally, if required, the structure was mirrored with respect to the x- or y-axis in order to get one helium atom in the first quadrant and the other two in the third and fourth quadrants. The corresponding normalized structures are plotted in (D)-(F). The structures are normalized to the largest pair distance, and subsequently located such that the two atoms with the largest pair distance get coordinates (-0.5,0) and (0.5,0) and the position of the third atom is plotted. The grey rectangle in (E) relates to structures that were cut during the reconstruction (see the supplementary materials for details). The linear color scale encodes the number of entries. The typical structure of the excited state of the He$_3$ is sketched in the upper right corner of (D).**

The geometry of the Efimov state is remarkably different from that of the ground state. Whereas the ground state corresponds to an almost randomly distributed cloud of particles (*24*), the excited Efimov state is dominated by configurations in which two atoms are close to each other and the third one further away, in the classically forbidden region of the two-body interaction potential. According to theory the average value of the smallest angle in the triangle is 18 degrees. This typical structure of the $^4$He$_3$ excited state is in accordance with the prediction of Nielsen and coworkers (*25*) and in line with a dimer-like pair model proposed by Hiyama and Kamimura (*28*), which explains the asymptotic behavior of the excited state of the He$_3$. More generally, the typical structure revealed by Figure 4 is an intrinsic property of all Efimov states with positive two-body scattering length. Comparison with Figure 1B shows that the geometrical structure of the real He$_3$ exited state is similar to the one of the hypothetical helium trimer with an infinite scattering length, though the amount of triangles with a small acute angle is reduced in the latter case (see also Figure S7 in the supplementary materials).

In conclusion, we have reported the observation of the elusive Efimov state of the $^4$He trimer by means of Coulomb explosion imaging of mass-selected clusters. We have found that the dominant structure of an Efimov state with positive two-body scattering length is a triangle with a relatively small acute angle. This has important consequences for how Efimov states are formed and possibly excited, as it is the geometry that defines the Franck-Condon overlap with continuum as well as with bound states. Having demonstrated the ability to experimentally image the quantum mechanical probability distribution of Efimov trimers, this work opens the door for quantitative studies of Efimov physics beyond the geometric scaling properties and energetics. Extensions of the imaging approach to the four-body sector appear feasible.

**Acknowledgments:** The experimental work was supported by a Reinhart Koselleck project of the Deutsche Forschungsgemeinschaft. The financial support of the Goethe-University Frankfurt am Main is gratefully acknowledged. D.B. acknowledges support by the US National Science Foundation through grant number PHY-1205443. Authors thank Dr. J. Williams for the proofreading of the manuscript. Raw data are archived at the Goethe-University Frankfurt am Main and are available upon request.

**Materials and Methods:** Figs. S1 to S7, References (29-31).

# Supplementary material
# Observation of the Efimov state of the helium trimer

Maksim Kunitski[1]*, Stefan Zeller[1], Jörg Voigtsberger[1], Anton Kalinin[1], Lothar Ph. H. Schmidt[1], Markus Schöffler[1], Achim Czasch[1], Wieland Schöllkopf[2], Robert E. Grisenti[1,3], Till Jahnke[1], Dörte Blume[4] and Reinhard Dörner[1]*

[1]Institut für Kernphysik, Goethe-Universität Frankfurt am Main, Max-von-Laue-Straße 1, 60438 Frankfurt am Main, Germany.

[2]Department of Molecular Physics, Fritz-Haber-Institut, Faradayweg 4-6, 14195 Berlin, Germany

[3]GSI Helmholtz Center for Heavy Ion Research, Planckstraße 1, 64291 Darmstadt, Germany

[4]Department of Physics and Astronomy, Washington State University, Pullman, WA 99164-2814, USA

## Content

## Trimer preparation and detection

The He clusters were prepared in a molecular beam under supersonic expansion of gaseous helium at a temperature of 8 K through a 5µm nozzle. The nozzle temperature was stabilized within +/- 0.02 K by a cryostat (Model 32B, Cryogenic Control Systems, Inc.) cooled with liquid helium. The cluster yields were tuned by varying the nozzle back pressure, which was stabilized within +/- 2 mbar below and +/-5 mbar above 1 bar (absolute values of pressure are used throughout the text). He trimers were selected from the molecular beam by means of matter wave diffraction using a transmission grating with a period of 100 nm (17,30,31); see Fig. S1 for a schematics. The selection allowed to get rid of an overwhelming fraction of He monomers that dominates in the molecular beam under all expansion conditions (28).

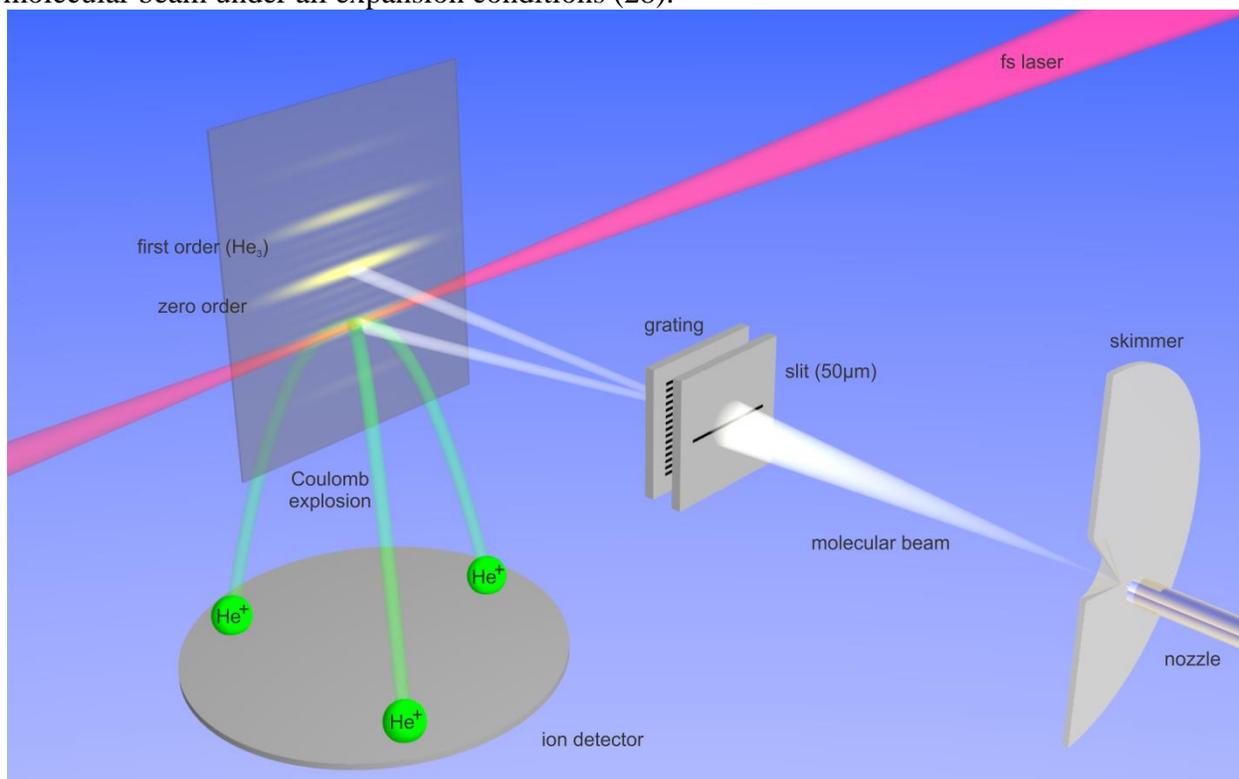

**Fig. S1. Selection of He trimers from the molecular beam by means of matter wave diffraction. The distances between the nozzle and skimmer, skimmer and slit, slit and grating, grating and ionization region were 15 mm, 332 mm, 30 mm and 491 mm, respectively.**

The ionization of all three atoms of the trimer by a strong ultra-short laser field (30 fs, 780 nm, $> 3 \cdot 10^{15}$ W/cm$^2$, Dragon KMLabs) led to Coulomb explosion (22) of the cluster. The momenta of the ions acquired during explosion were measured by cold target recoil ion momentum spectroscopy (COLTRIMS) (23). In the COLTRIMS spectrometer a homogenous electric field of 3.12 V/cm guided the ions onto a time and position sensitive micro-channel plate detector with hexagonal delay-line position readout (24) and an active area of 80 mm. The detector was placed 38 mm away from the laser focus which resulted in a 4π collection solid angle for the atomic ions with an energy of up to 3 eV.

## Separation of the ground and excited states in momentum space

As the ground and excited state wave functions are very different in their spatial extents, their patterns in momentum space after Coulomb explosion can be fairly well separated in a two-dimensional plot, where the Kinetic Energy Release (KER) (see also equation S3 below) is plotted versus the magnitude of the smallest momentum among the three particles (Fig S2). To cross check the separation quality of this two-dimensional filter we have simulated the Coulomb explosion for theoretical structures of the ground and excited states (Fig. S2 A,B). This simulation showed that our filter misses about 20% of the excited state structures. This is the price we have to pay for keeping the contamination of the excited state with ground state structures low at about 0.03% (with respect to the amount of the ground state). Since the ratio of the ground to excited state in the experiment was about 100 under best expansion conditions (8K, 330 mbar), the contamination of the excited state after the filter is estimated to be around 3%.

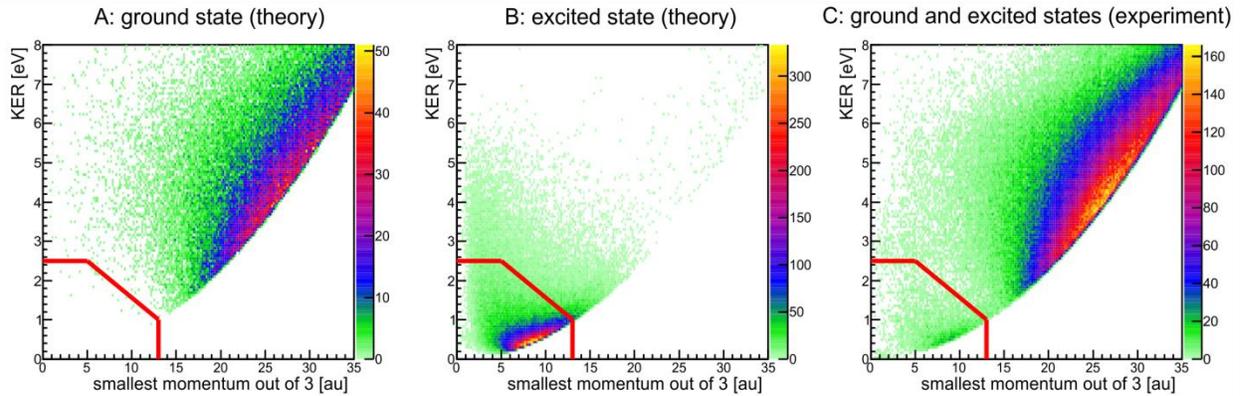

**Fig. S2. Kinetic energy release vs. the smallest momentum after Coulomb explosion for the theoretical ground state (A) and excited state (B) of $^4$He$_3$ as well as for the experimentally measured mixture of the ground and excited states (C, nozzle: 8 K, 330 mbar). The red boundary defines the filter, which was used in order to separate the excited state of $^4$He$_3$ from the ground state.**

## Reconstruction of structures with the look-up table approach

The highly nonlinear relation between the initial spatial geometry and the momenta after Coulomb explosion, combined with the small initial speeds of the ions due to ionization (see below for details), render an iterative procedure to invert the Coulomb explosion impracticable. We therefore devised a look-up table approach for the inversion. Following the suggestion of Dalitz (32) we define the following coordinates in real (coordinate) and momentum space (see Fig. S3):
Coordinate space:

$$X^{Dalitz}_{coord} = \frac{|\vec{r}_2|^2 - |\vec{r}_3|^2}{\sqrt{3} \cdot \sum_{i=1}^{3}|\vec{r}_i|^2}, \quad Y^{Dalitz}_{coord} = \frac{|\vec{r}_1|^2}{\sum_{i=1}^{3}|\vec{r}_i|^2} - \frac{1}{3}, \quad (S1)$$

where $\vec{r}_i$ denotes the position vector of the $i$th atom of the trimer with respect to the center-of-mass.
Momentum space:

$$X^{Dalitz}_{momentum} = \frac{|\vec{p}_2|^2 - |\vec{p}_3|^2}{\sqrt{3} \cdot \sum_{i=1}^{3}|\vec{p}_i|^2}, \quad Y^{Dalitz}_{momentum} = \frac{|\vec{p}_1|^2}{\sum_{i=1}^{3}|\vec{p}_i|^2} - \frac{1}{3}, \quad (S2)$$

where $\vec{p}_i$ denotes the momentum vector of the $i$th ion after Coulomb explosion.

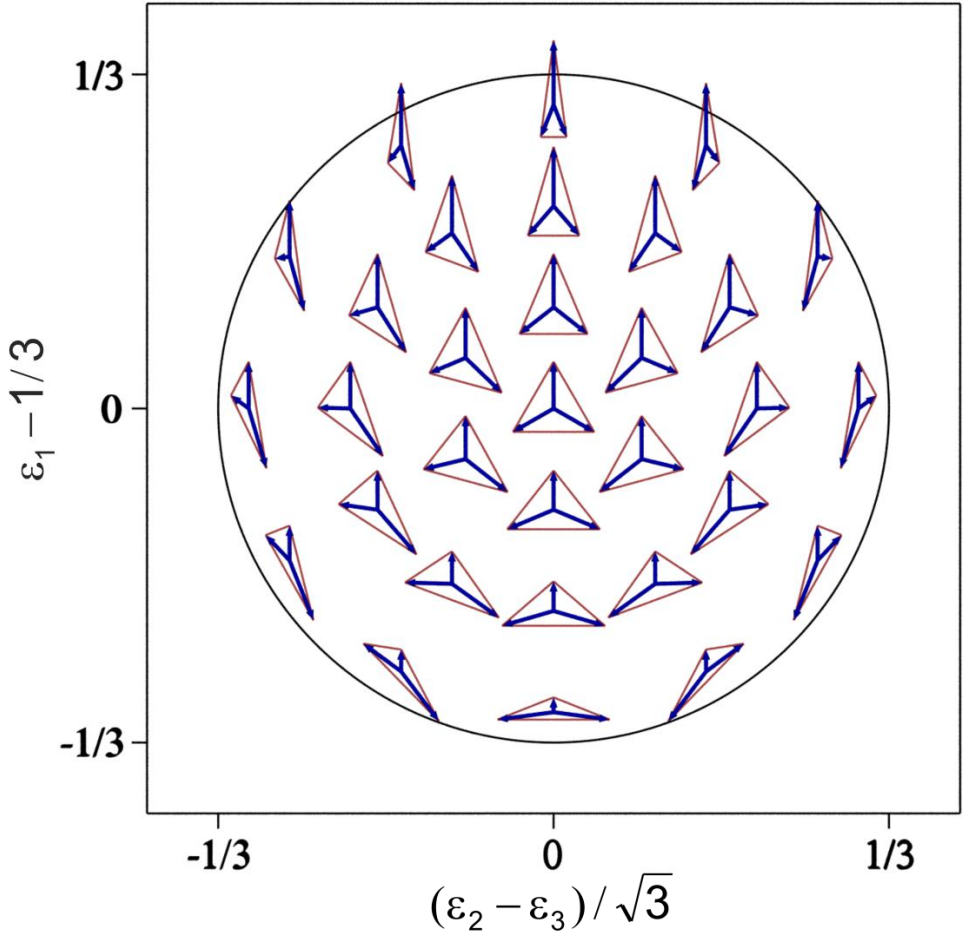

**Fig. S3. The Dalitz plot.** The $\varepsilon_i$ are defined through $\varepsilon_i = |\vec{p}_i|^2 / \sum_i |\vec{p}_i|^2$. **The blue arrows show the vectors $\vec{p}_i$ for various representative triangles in the center-of-mass frame, i.e. such that $\sum_i \vec{p}_i = 0$. The vector $\vec{p}_1$ points always upwards. The $\vec{p}_i$ vectors can either be momentum vectors (momentum Dalitz plot) or coordinate vectors defined with respect to the center-of-mass origin of the triangle (coordinate Dalitz plot).**

For 1000x1000 different structures characterized by $X^{Dalitz}_{coord}, Y^{Dalitz}_{coord}$ we calculated the final momentum vectors and subsequently $X^{Dalitz}_{momentum}, Y^{Dalitz}_{momentum}$ using Newton's equations of motion. The trajectories were launched six times with different randomly generated small initial momenta. The distribution of these initial momenta was taken from the measured momentum distribution of the singly charged helium ion. These initial momenta originate from the recoil of the emitted electron. We checked that the initial values of the momenta do not alter the results significantly. This procedure allowed us to build a look-up table, where each 2D point ($X^{Dalitz}_{momentum}, Y^{Dalitz}_{momentum}$) in Dalitz momentum space (out of 1000x1000) was connected to several 2D points ($X^{Dalitz}_{coord}, Y^{Dalitz}_{coord}$) in Dalitz coordinate space. This lookup table was then used to obtain ten different (randomly chosen) coordinates ($X^{Dalitz}_{coord}, Y^{Dalitz}_{coord}$) corresponding to measured momenta ($X^{Dalitz}_{momentum}, Y^{Dalitz}_{momentum}$). Finally, the absolute structure for each event was obtained from Dalitz coordinate space ($X^{Dalitz}_{coord}, Y^{Dalitz}_{coord}$) utilizing the measured kinetic energy release (KER) of this event and the following relation between KER and the pair distances $R_{ij}$ in the trimer (in atomic units):

$$KER = 1/R_{12} + 1/R_{13} + 1/R_{23} \quad (S3)$$

## Reconstruction check

In order to check our structure reconstruction procedure we calculated the momenta acquired during Coulomb explosion of the theoretical excited state structures. The initial momenta of the ions prior to explosion were randomly chosen within the experimentally measured distribution (see also discussion above). Using the calculated momenta, we construct the momentum Dalitz plot (see Fig. S4 B) and retrieve the corresponding trimer structure with our look-up table (see Fig. S4 C and D). Our procedure reliably retrieved the initial structure except for a very small portion of the configuration space that is located on the outer radius of the Dalitz plot (marked with L in Fig. S4 C). The physics behind this failure is that there exist two spatial configurations for which Coulomb explosion leads to one very low energy ion in conjunction with two almost back-to-back emitted ions. One configuration corresponds to a nearly linear geometry, where the third ion sits in between the other two ions (see the three regions marked by L in Fig. S4 C). The second configuration corresponds to a "2+1" geometry, where two ions are close to each other with the third one further away (see the three regions marked by R in Fig. S4 C). According to the theory of Efimov trimers, the nearly linear configuration is almost absent. To avoid artefacts we therefore rejected all nearly linear geometries during our look-up table reconstruction. These configurations are located in the very narrow region of the structure space colored in grey in Fig. S4 D. Since there are typically several entries in our look-up table for each ($X_{momentum}^{Dalitz}, Y_{momentum}^{Dalitz}$) coordinate, we use, if the reconstruction generates an event in the rejected area, an alternative Dalitz coordinate configuration that is associated with the same ($X_{momentum}^{Dalitz}, Y_{momentum}^{Dalitz}$). Nevertherless about 10% of measured momenta could not be inverted into the structure because of the absence of valid entries in the look-up table.

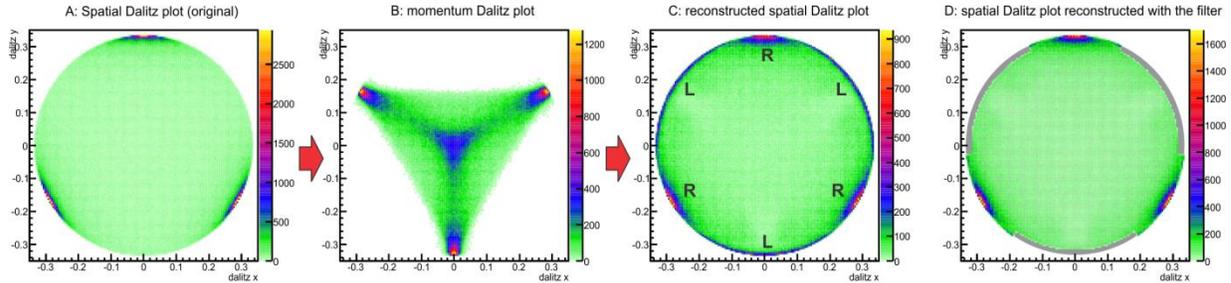

**Fig. S4. Test of the look-up table based reconstruction algorithm using theoretical structures of the Efimov trimer. A) Dalitz plot of theoretical structures in coordinate space (equation S1). B) Dalitz plot in momentum space (equation S2) obtained by simulating the Coulomb explosion of the coordinate structures shown in A. C) Dalitz plot in coordinate space, obtained by reconstruction of the coordinate structures from the momentum structures shown in B using our look-up table approach without rejection of nearly linear configurations. Note the artefacts on the outer ring of the plot in the vicinity of the regions marked by "L". D) the same as C) but with replacing configurations in the grey areas with other configurations ($X_{coord}^{Dalitz}, Y_{coord}^{Dalitz}$) that originate from the same ($X_{momentum}^{Dalitz}, Y_{momentum}^{Dalitz}$) coordinates.**

The theoretical pair-distance distribution of the excited He₃ (corresponding to the coordinate Dalitz plot shown in Fig. S4 A) is plotted in Fig. S5 (blue curve). The reconstructed pair-distance distribution (related to the coordinate Dalitz plot shown in Fig. S4 D) is shown in Fig. S5 in red.

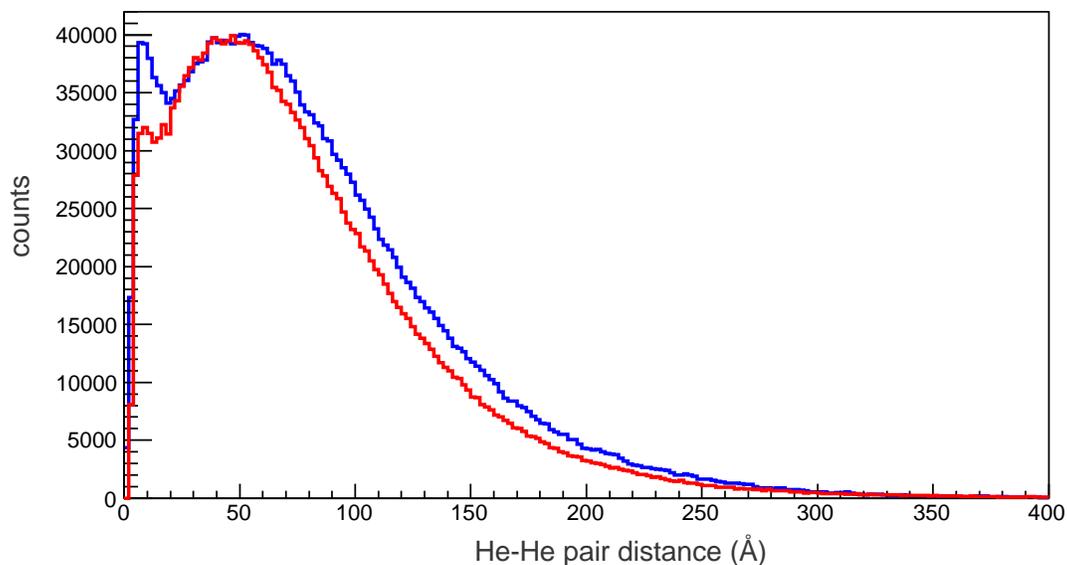

**Fig. S5 The reconstruction quality check. The blue curve is the pair-distance distribution for the theoretical excited state structures of He$_3$. The reconstruction of these structures from the momentum space Dalitz plot, obtained by simulating the Coulomb explosion, yields the red distribution. The lower number of entries of the reconstructed distribution (red) with respect to the original one (blue) is due to the event loss (about 10%) during reconstruction.**

## Binding energy fits

To determine the binding energy of the excited helium trimer we have fitted the corresponding pair-distance distribution with the exponential decay function (see equation 2 of the main article). As a result the energy difference $\Delta B$ between the binding energies of the excited trimer and of the ground state dimer was found to be 0.98±0.2 mK (Fig. S6). The error has been estimated very conservatively including the statistical quality of our fit and the systematic errors due to our reconstruction procedure.

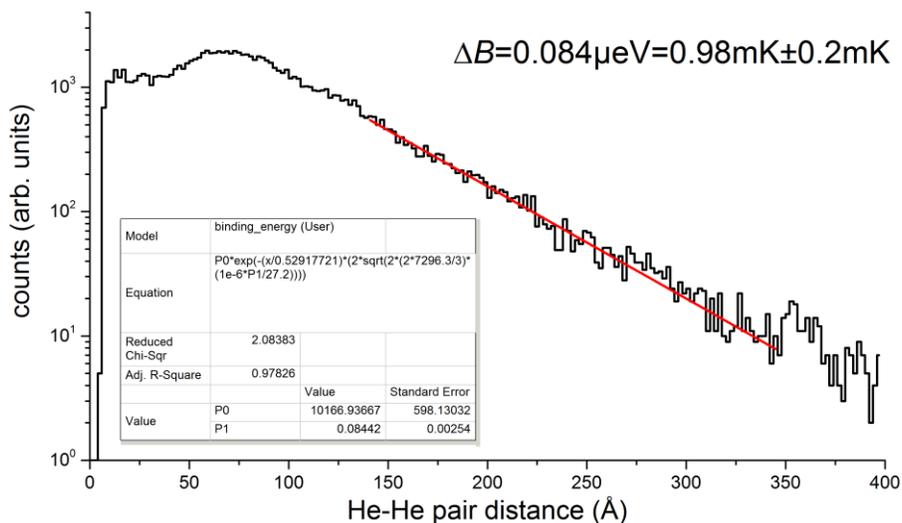

**Fig. S6. Fit of the falling edge of the reconstructed pair distance distribution of excited He$_3$ with an exponential decay function (see equation 2 of the main article). $\Delta B$ is the difference between the binding energies of the excited state of the trimer and the ground state of the dimer.**

## The error analysis

The error bars in Figs. 2, 3 correspond to statistical errors, which have been estimated according to the Poisson distribution. Namely, the mean value for each bin in histograms as well as the variance *Var* equals to the number of entries $N$ in the bin. For large $N$ the Poisson distribution accurately resembles the Gaussian distribution with the mean value $N$ and the standard deviation $\sigma = \sqrt{Var} = \sqrt{N}$. Thus, the confidence intervals of 68% and 95% correspond to $\pm\sigma$ ($\pm\sqrt{N}$) and $\pm2\sigma$ ($\pm2\sqrt{N}$), respectively. The standard deviation for the sum of two independent random variables (difference distributions in Figs. 2, 3) was calculated according to the following equation:

$$\sigma_{aX+bY} = \sqrt{(a \cdot \sigma_X)^2 + (b \cdot \sigma_Y)^2}$$

## Theory

The He trimer potential energy surface is written as a sum of two-body potentials $V_{He-He}$ (11); the two-body potential accounts for retardation effects. To solve the time-independent Schrödinger equation, we separate off the center-of-mass degrees of freedom. Restricting ourselves to eigen states with vanishing relative three-body angular momentum, the relative Schrödinger equation depends on three relative coordinates, which are taken to be one radial coordinate (the hyperradius) and two angles. Following Ref. (20), the relative Schrödinger equation is solved in a two-step process. First we solve a two-dimensional differential equation in the angular degrees of freedom. The resulting eigenvalues and coupling matrix elements then enter into a coupled set of hyperradial equations. Our calculation employs up to 12 channels and yields ground and excited state energies that are, within errorbars, consistent with those reported in Ref. (12).

The pair distance and KER distributions for the ground and excited trimer states were calculated from the square of the respective three-dimensional wave functions. For the KER distribution, e.g., the quantity $1/R_{12}+1/R_{13}+1/R_{23}$ was averaged over the three-dimensional density and binned. The theoretical structures shown in Figs. 1, 4, S2 and S4 were generated by selecting around 80,000 geometries according to the three-dimensional probability densities of the ground and excited states.

## Comparison of the hypothetical He₃ structures for different scattering lengths

In order to compare structures of the excited state of the hypothetical helium trimer for different scattering lengths, Fig. S7 shows the distribution of the ratio of the shortest interparticle distance to the longest one ($R_{min}/R_{max}$) for $a$=-453 Å (just before the excited state disappears in the three-particle continuum), $a$=∞ and $a$=90.4 Å (true helium trimer). The ground state (GS) distribution of the true He₃ is shown for comparison. It is seen that the typical excited state structure of the true He₃ is where two atoms are close to each other with the third one being far away (red distribution). In this case, the distribution peaks at $R_{min}/R_{max}$ equal to around 0.1. In contrast, such structures are almost absent in the ground state of the true He₃ (black distribution). By tuning the He-He potential the artificial He₃ trimers with different scattering lengths can be created. By going from positive to negative scattering lengths the amount of the excited state structures with a relatively small acute angle decreases (red > green > blue distributions). However, even on the negative scattering length side ($a$=-453 Å; further weakening of the He-He potential, and correspondingly a less negative scattering length, leads to the disappearing of the excited state in the three-particle continuum), the amount of these structures is rather high (blue distribution).

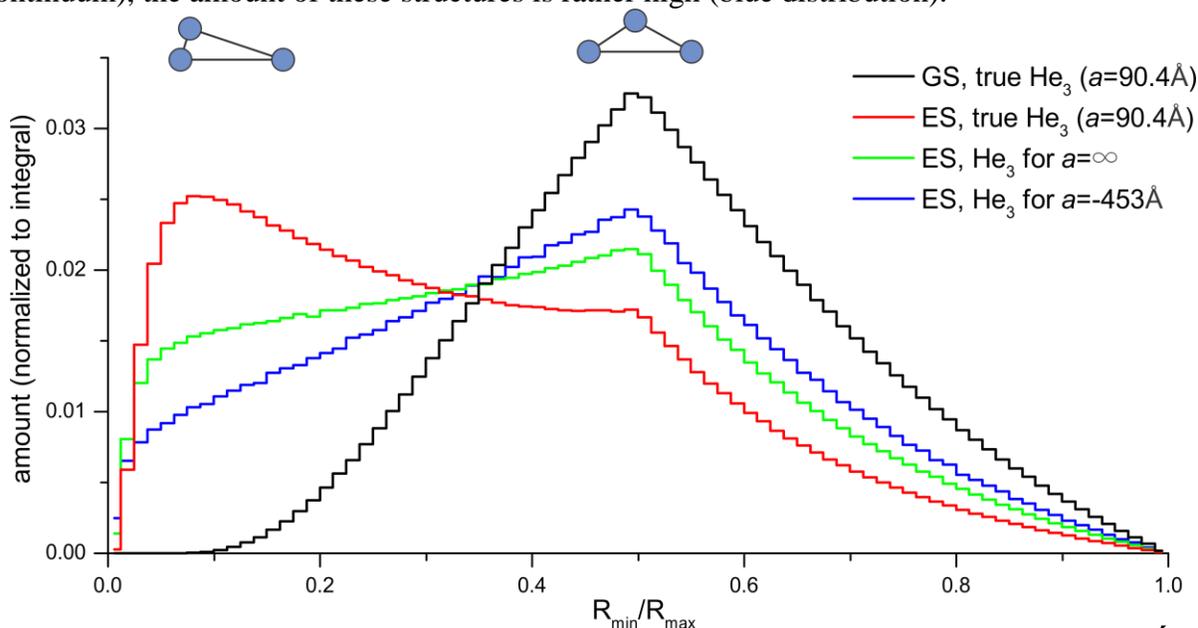

**Fig. S7. Comparison of structures of the true helium trimers (scattering length $a$=90.4 Å) and of two hypothetical helium trimers with scattering lengths $a$=-453 Å and $a$=∞. GS and ES stand for the ground state and the excited state, respectively. The distributions are obtained from our full quantum mechanical calculations.**